\documentclass[preprint,onecolumn,aps,pre,eqsecnum,showpacs,showkeys,a4paper]{revtex4-1}
\usepackage[T1]{fontenc}
\usepackage[latin9]{inputenc}
\usepackage{color}
\definecolor{note_fontcolor}{rgb}{0.800781, 0.800781, 0.800781}
\usepackage[english]{babel}
\usepackage{amsmath}
\usepackage{amssymb}
\usepackage{graphicx}
\usepackage[unicode=true,
 bookmarks=true,bookmarksnumbered=true,bookmarksopen=true,bookmarksopenlevel=1,
 breaklinks=true,pdfborder={0 0 0},backref=false,colorlinks=true]
 {hyperref}
\hypersetup{
 pdfauthor={AINS}}

\makeatletter

\newenvironment{lyxgreyedout}
  {\textcolor{note_fontcolor}\bgroup\ignorespaces}
  {\ignorespacesafterend\egroup}

\numberwithin{equation}{section}
\numberwithin{figure}{section}
\numberwithin{table}{section}

\usepackage{graphicx}

\DeclareGraphicsExtensions{.png,.pdf,.eps}
\graphicspath{{pictures/}}

\@ifundefined{showcaptionsetup}{}{%
 \PassOptionsToPackage{caption=false}{subfig}}
\usepackage{subfig}
\makeatother

\begin{document}

\title{The Quantum Sweeper Effect\vspace*{\bigskipamount}
}

\author{Gerhard \surname{Grössing}\textsuperscript{}}

\email[E-mail: ]{ains@chello.at}

\homepage[Visit: ]{http://www.nonlinearstudies.at/}

\author{Siegfried \surname{Fussy}\textsuperscript{}}

\email[E-mail: ]{ains@chello.at}

\homepage[Visit: ]{http://www.nonlinearstudies.at/}

\author{Johannes \surname{Mesa Pascasio}\textsuperscript{}}

\email[E-mail: ]{ains@chello.at}

\homepage[Visit: ]{http://www.nonlinearstudies.at/}

\author{Herbert \surname{Schwabl}\textsuperscript{}}

\email[E-mail: ]{ains@chello.at}

\homepage[Visit: ]{http://www.nonlinearstudies.at/}

\affiliation{\textsuperscript{}Austrian Institute for Nonlinear Studies, Akademiehof\\
 Friedrichstr.~10, 1010 Vienna, Austria\\
\vspace*{1cm}
}

\date{\today}
\begin{abstract}
We show that during stochastic beam attenuation in double slit experiments,
there appear unexpected new effects for transmission factors below
$a\lesssim10^{-4}$, which can eventually be observed with the aid
of weak measurement techniques. These are denoted as \textsl{quantum
sweeper} effects, which are characterized by the bunching together
of low counting rate particles within very narrow spatial domains.
We employ a ``superclassical'' modeling procedure which we have
previously shown to produce predictions identical with those of standard
quantum theory. Thus it is demonstrated that in reaching down to ever
weaker channel intensities, the nonlinear nature of the probability
density currents becomes ever more important.  We finally show that
the resulting unexpected effects nevertheless implicitly also exist
in standard quantum mechanics.%
\begin{lyxgreyedout}
\noindent \global\long\def\VEC#1{\mathbf{#1}}
\global\long\def\d{\,\mathrm{d}}
\global\long\def\e{{\rm e}}
\global\long\def\meant#1{\left<#1\right>}
\global\long\def\meanx#1{\overline{#1}}
\global\long\def\mpbracket{\ensuremath{\genfrac{}{}{0pt}{1}{-}{\scriptstyle (\kern-1pt +\kern-1pt )}}}
\global\long\def\pmbracket{\ensuremath{\genfrac{}{}{0pt}{1}{+}{\scriptstyle (\kern-1pt -\kern-1pt )}}}
\global\long\def\p{\partial}
\end{lyxgreyedout}

\end{abstract}
\maketitle

\section{Introduction\label{sec:Introduction}}

In his criticism of David Bohm's causal interpretation of the quantum
mechanical formalism, Isidor Rabi made the following statement in
the 1950ies which is still shared by quite some researchers today:
``I do not see how the causal interpretation gives us any line to
work on other than the use of the concepts of quantum theory. Every
time a concept of quantum theory comes along, you can say yes, it
would do the same thing as this in the causal interpretation. But
I would like to see a situation where the thing turns around, when
you predict something and we say, yes, the quantum theory can do it
too.'' \cite{FreireJr..2005science}

Although doubtlessly the project of a causal interpretation á la de~Broglie--Bohm
has gained momentum in recent years, with many of its results exhibiting
more detailed illustrations via particle trajectories, in fact no
experimental prediction unknown to orthodox quantum theory has yet
arisen from this approach.

However, in the present paper we discuss a finding based on a causal
view of quantum mechanics that amounts to a new effect which has so
far eluded orthodox quantum mechanics. Specifically, we recently reported
theoretical results from extreme beam attenuation techniques in double-slit
experiments whose phenomenology is described by what we term the \emph{quantum
sweeper effect}. \cite{Groessing.2014attenuation} The discovery of
this effect is due to our causal subquantum model for quantum systems,
whose central features can be shown to exactly match with the de~Broglie--Bohm
theory \cite{Groessing.2015implications}.

In Chapter 2 of this paper, we firstly present the quantum sweeper
effect in said subquantum scenario. Chapter 3, then, discusses the
consequences of the mentioned beam attenuation techniques in purely
quantum mechanical language, thus showing that the latter does provide
statements in line with, but not as detailed as, the effect described
in subquantum terms.

Finally, in Chapter 4 we shall discuss the significance of the sweeper
effect. Essentially, we show that in employing ever weaker channel
intensities, nonlinear effects become ever more important, which are
generally not considered -- although implicitly present -- in ordinary
quantum mechanics. The latter are a crucial characteristic of subquantum
models as the one developed by our group, with experimental tests
becoming feasible through the use of weak measurement techniques.
Consequences are also discussed with respect to the meaning of the
complementarity principle in the light of our new findings.

Before we discuss the sweeper effect, however, some preliminaries
are necessary in order to provide the appropriate context. To begin
with, we recall from~\cite{Rauch.1984static,Rauch.1990low-contrast}
that a beam chopper was employed as a deterministic absorber in one
arm of a two-armed interferometer, whereas for stochastic absorption
semitransparent foils of various materials were used. Despite the
net effect of the same percentage of neutrons being attenuated, the
quantum mechanical formalism predicts the following different behaviors
for the two cases. Introducing the \emph{transmission factor} $a$
as the beam's transmission probability, in the case of a (deterministic)
chopper wheel it is given by the temporal open-to-closed ratio, $a=\frac{t_{\mathrm{open}}}{t_{\mathrm{open}}+t_{\mathrm{closed}}}$
, whereas for a (stochastic) semitransparent material defined by its
absorption cross section, it is simply the relation of the intensity
$I$ with absorption compared to the intensity $I_{0}$ without, i.e.\ $a=I/I_{0}$.
Thus the beam modulation behind the interferometer is obtained in
the following two forms. For the deterministic chopper system the
intensity is, with $\varphi$ denoting the phase difference, given
by
\begin{equation}
I\propto\left(1-a\right)\left|\varPsi_{1}\right|^{2}+a\left|\varPsi_{1}+\varPsi_{2}\right|^{2}\propto1+a+2a\cos\varphi,\label{eq:sw.1}
\end{equation}
whereas for stochastic beam attenuation with the semitransparent material
it is
\begin{equation}
I\propto\left|\varPsi_{1}+\varPsi_{2}\right|^{2}\propto1+a+2\sqrt{a}\cos\varphi.\label{eq:sw.2}
\end{equation}
In other words, although the same number of neutrons is observed in
both cases, in the first one the contrast of the interference pattern
is proportional to $a$, whereas in the second case it is proportional
to $\sqrt{a}$.

In \cite{Groessing.2014attenuation} we accounted for the just described
attenuation effects, and we chose the usual double slit scenario,
primarily because this turned out to be very useful when discussing
more extreme intensity hybrids (i.e., combinations of high and very
low transmission factors). Moreover, in this way we could make simple
use of the essentials of our model which are employed in our explanation
of the beam attenuation phenomena.

Throughout the last years we have developed an approach to quantum
mechanics within the scope of theories on ``Emergent Quantum Mechanics''.
(For the proceedings of the first two international conferences devoted
to this subject, see~\cite{Groessing.2012emerqum11-book,Groessing.2014emqm13-book}.)
Essentially, we consider the quantum as a complex dynamical phenomenon,
rather than as representing some ultimate-level phenomenon in terms
of, e.g., pure formalism, wave mechanics, or strictly particle physics
only. Our assumption is that a particle of energy $E=\hbar\omega$
is actually an oscillator of angular frequency $\omega$ phase-locked
with the zero-point oscillations of the surrounding environment, the
latter of which containing both regular undulatory and fluctuating
components and being constrained by the boundary conditions of the
experimental setup via the emergence of standing waves. In other words,
the particle in this approach is an off-equilibrium steady-state maintained\ by
the throughput of zero-point energy from its vacuum surroundings.
This is in close analogy to the bouncing/walking droplets in the experiments
of Couder and Fort's group~\cite{Fort.2010path-memory,Couder.2006single-particle,Couder.2012probabilities},
which in many respects can serve as a classical prototype guiding
our intuition. However, we denote our whole ansatz as ``superclassical''~\cite{Groessing.2014emqm13-book},
because it connects the classical physics at vastly different scales,
i.e.\ the ordinary classical one and an assumed subquantum one, with
``new'' effects emergent on intermediate scales, which we have come
to know and describe as quantum ones.

In fact, we have succeeded in reproducing a number of quantum mechanical
results with our superclassical model, i.e.\ without any use of the
quantum mechanical formalism, like states, wave functions, \textit{et
cetera}. Note, moreover, that a Gaussian emerging from, say, a single
slit with rounded edges (so as to avoid diffraction effects) is in
our model the result of statistically collecting the effects of the
aleatory bouncing of our particle oscillator. Rather, the Gaussian
stands for the statistical mean of the ``excitation'' (or ``heating
up'') of the medium within the confines of the slit, and later, as
the bouncer/walker progresses, further away from it. We have described
this in terms of a thermal environment that represents stored kinetic
energy in the vacuum and that is responsible for where the particle
is being guided to. For example, consider particle propagation coming
out from a Gaussian slit. For a particle exactly at the center of
the Gaussian, the diffusive momentum contributions from the heated
up environment will on average cancel each other for symmetry reasons.
However, the further off the particle is from that center, the stronger
the symmetry will be broken, thus resulting in a position-dependent
net acceleration or deceleration, respectively~-- in effect, resulting
in the decay of the wave packet. (For a detailed analysis, see~\cite{Groessing.2010emergence}.)
In other words, due to wave-like diffusive propagations originating
from the particle's bounces to stir up the medium of the vacuum, particle
paths can be influenced by the agitations of the vacuum even in places
where no other particle is around. 

As already mentioned, we have shown that the spreading of a wave packet
can be exactly described by combining the forward (convective) with
the orthogonal diffusive velocity fields. The latter fulfill the condition
of being unbiased w.r.t.\ the convective velocities, i.e.~the orthogonality
relation for the \textit{averaged} velocities derived in~\cite{Groessing.2010emergence}
is $\VEC{\overline{vu}}=0$, since any fluctuations $\VEC u=\delta\left(\nabla S/m\right)$
are shifts along the surfaces of action $\mathit{S=\mathrm{\mathrm{const}}.}$
Moreover, the fluctuations can be directed towards the left or towards
the right from the mean (i.e.\ from the Ehrenfest trajectory), which
leads us to introduce the notations $\mathbf{u}_{i\mathrm{L}}$ and
$\mathbf{u}{}_{i\mathrm{R}}$, respectively.

Reducing the general case discussed in~\cite{Fussy.2014multislit}
to the double-slit case, one notes for the first and second channels
the emergent velocity vectors $\VEC v_{\mathrm{1(2)}},\VEC u_{\mathrm{1(2)R}},$
and $\VEC u_{\mathrm{1(2)L}}$, together with associated amplitudes
$R_{1(2)}$, respectively. In order to completely accommodate the
totality of the system of currents present, one obtains a local wave
intensity for any velocity component, e.g.\ for $\VEC v_{\mathrm{1}}$,
by the pairwise projection on the unit vector $\VEC{\hat{v}}_{1}$
weighted by $R_{1}$ of the totality of all amplitude weighted unit
velocity vectors being operative at $\mathrm{(}\VEC x,t)$:
\begin{equation}
P(\VEC v_{\mathrm{1}})=R_{1}\VEC{\hat{v}}_{1}\cdot(\VEC{\hat{v}}_{1}R_{1}+\VEC{\hat{u}_{\mathrm{1R}}\mathrm{\mathit{R_{\mathrm{1}}}}}+\VEC{\hat{u}_{\mathrm{1L}}\mathrm{\mathit{R_{\mathrm{1}}}}}+\VEC{\hat{v}}_{2}R_{2}+\VEC{\hat{u}_{\mathrm{2R}}\mathit{R_{\mathrm{2}}}}+\VEC{\hat{u}}_{\mathrm{2L}}R_{2}).\label{eq:sw.3}
\end{equation}
The local current attributed to each velocity component is defined
as the corresponding ``local'' intensity-weighted velocity, e.g.\ for
$\VEC v_{1}$ it is given as $\VEC J\mathrm{(}\VEC v_{\mathrm{1}})=\VEC v_{\mathrm{1}}P(\VEC v_{\mathrm{1}})=\VEC v_{\mathrm{1}}\left(R_{1}^{2}+R_{1}R_{2}\cos\varphi\right)$.
The local intensity of a partial current is dependent on all other
currents, and the total current itself is composed of all partial
components, thus constituting a representation of what we call \emph{relational
causality}. After a short calculation, the total current turns out
as $\VEC J_{\mathrm{tot}}=\VEC v_{\mathrm{1}}P(\VEC v_{\mathrm{1}})+\VEC u_{1R}P\mathrm{(}\VEC u_{1\mathrm{R}}\mathrm{)}+\VEC u_{1\mathrm{L}}P\mathrm{(}\VEC u_{\mathrm{1\mathrm{L}}}\mathrm{)}+\VEC v_{\mathrm{2}}P(\VEC v_{\mathrm{2}})+\VEC u_{2R}P\mathrm{(}\VEC u_{2\mathrm{R}}\mathrm{)}+\VEC u_{\mathrm{2L}}P\mathrm{(}\VEC u_{2\mathrm{L}}\mathrm{)}$,
which, by identifying the resulting diffusive velocities $\VEC u_{i\mathrm{\mathrm{R}}}-\VEC u_{i\mathrm{L}}$
with the effective diffusive velocities $\VEC u_{i}$ for each channel,
finally leads to
\begin{equation}
\VEC J_{\mathrm{tot}}=R_{1}^{2}\VEC v_{\mathrm{1}}+R_{2}^{2}\VEC v_{\mathrm{2}}+R_{1}R_{2}\left(\VEC v_{\mathrm{1}}+\VEC v_{2}\right)\cos\varphi+R_{1}R_{2}\left(\VEC u_{1}-\VEC u_{2}\right)\sin\varphi.\label{eq:sw.10}
\end{equation}
The trajectories or streamlines, respectively, are given by
\begin{equation}
\VEC{\dot{x}}=\VEC v_{\mathrm{tot}}=\frac{\VEC J_{\mathrm{tot}}}{P_{\mathrm{tot}}}\thinspace.\label{eq:sw.11}
\end{equation}
As first shown in~\cite{Groessing.2012doubleslit}, by re-inserting
the expressions for convective and diffusive velocities, respectively,
i.e.\ 
\begin{equation}
\VEC v_{i}=\frac{\nabla S_{i}}{m},\quad\textrm{ and }\quad\VEC u_{i}=-\frac{\hbar}{m}\frac{\nabla R_{i}}{R_{i}}\,,\label{eq:sw.12}
\end{equation}
one immediately identifies Eq.~(\ref{eq:sw.11}) with the Bohmian
guidance equation and Eq.~(\ref{eq:sw.10}) with the quantum mechanical
pendant for the probability density current~\cite{Sanz.2008trajectory}.
As we have shown also the latter identity, we are assured that our
results are the same as those of standard quantum mechanics -- provided,
of course, that generally (i.e.\ in the standard quantum as well
as in our ansatz) the idealization of using Gaussians or similar regular
distribution functions is applicable for the high degrees of attenuation
studied here. However, as in our model we can also make use of the
velocity field to plot the averaged particle trajectories, we can
in principle provide a more detailed picture, in similar ways to the
Bohmian one, but still not relying on any quantum mechanical tool
like a wave function, for example.

In interpreting their results of the beam attenuation experiments
with neutrons, Rauch~\emph{et\,al.} found evidence in support of
the complementarity principle. That is, the more pronounced the visibility
of the interference fringes, the less which-path knowledge one can
have of the particle propagation, and \emph{vice versa}: the higher
the probability is for a particle to take a path through one certain
slit, the less visible the interference pattern becomes. This was
in fact confirmed in the above-mentioned neutron interferometry experiments,
albeit to a lesser degree for very low counting rates. In particular,
the authors of~\cite{Rauch.1984static,Rauch.1990low-contrast} often
use expressions such as the ``particle-like'' or the ``wave-like''
nature of the quantum system studied, depending on whether which-path
information or interference effects are dominant, respectively. While
this is all correct as far as the mentioned papers are concerned,
an extrapolation of the use of ``particle-like'' or ``wave-like''
attributed to more extreme intensity hybrids is not guaranteed. In
fact, we shall show below a particular effect which undermines said
dichotomy of ``particle-like'' and ``wave-like'' features, thereby
calling for an improved, more general analysis of possible relationships
between particle and wave features.

\section{The quantum sweeper effect during stochastic beam attenuation in
the double slit and its superclassical modeling\label{sec:Sweeper subquantum}}

Let us start with a discussion of stochastic attenuation using a coherent
beam in a double-slit experiment. With the intensity distribution
being recorded on a screen, we are going to discuss a particular effect
of the attenuation of one of the two emerging Gaussians at very small
transmission factors. With the appropriate filtering of the particles
going through one of the two slits, the recorded probability density
on a screen in the surroundings of the experiment will appear differently
than what one would normally expect. That is, even if one had a low
beam intensity coming from one slit, one would expect the following
scenario according to the usual quantum mechanical heuristics: The
interference pattern would more and more become asymmetric in the
sense that the contributions from the fully open slit would become
dominant until such a low counting rate from the attenuated slit is
arrived at that essentially one would have a one-slit distribution
of recorded particles on the screen. 

\begin{figure*}
\begin{centering}
\subfloat[$a=10^{-1}$\label{fig:sw.2a}]{\protect\begin{centering}
\protect\includegraphics[width=0.31\textwidth]{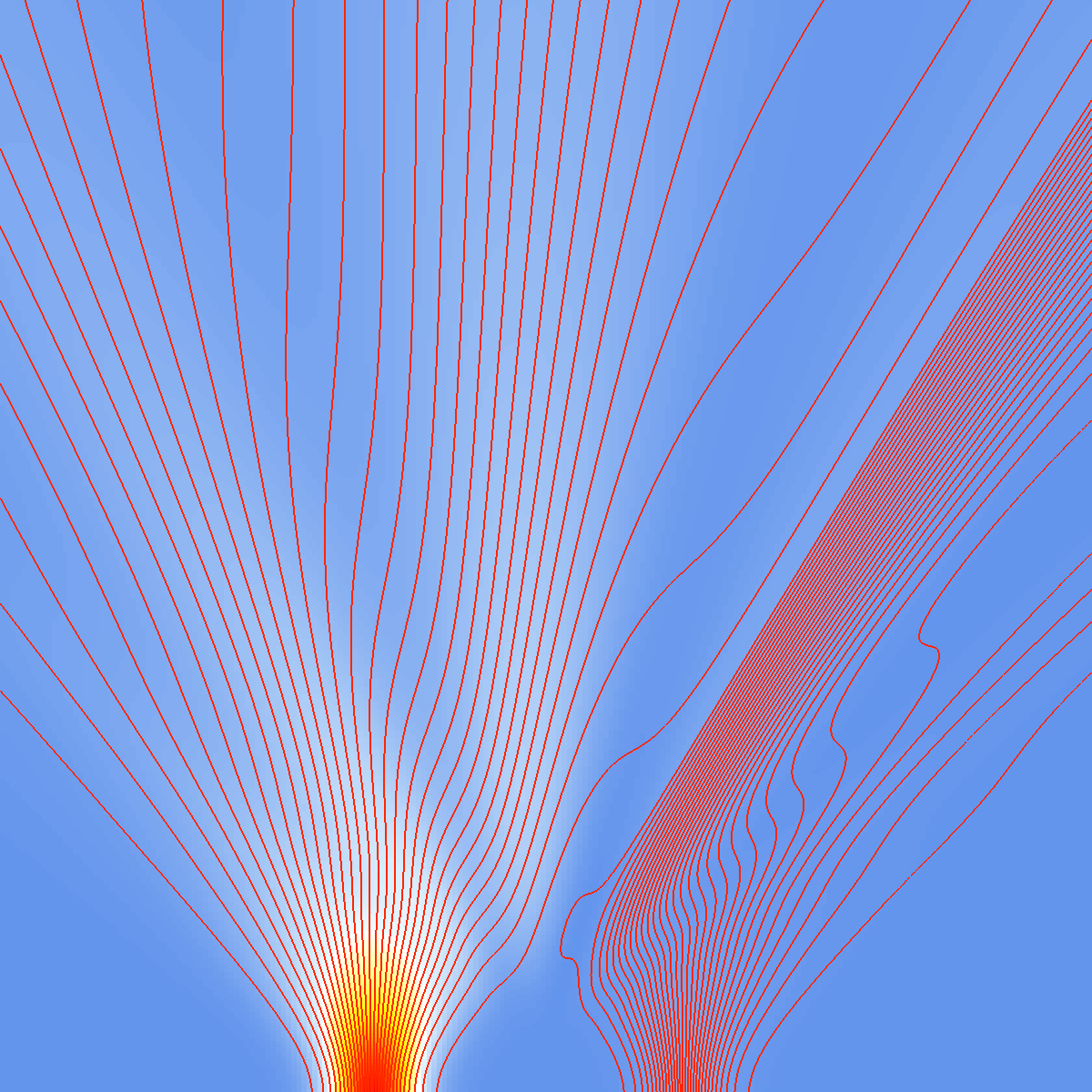} \protect
\par\end{centering}

}\,\subfloat[$a=10^{-4}$\label{fig:sw.2b-1}]{\protect\begin{centering}
\protect\includegraphics[width=0.31\textwidth]{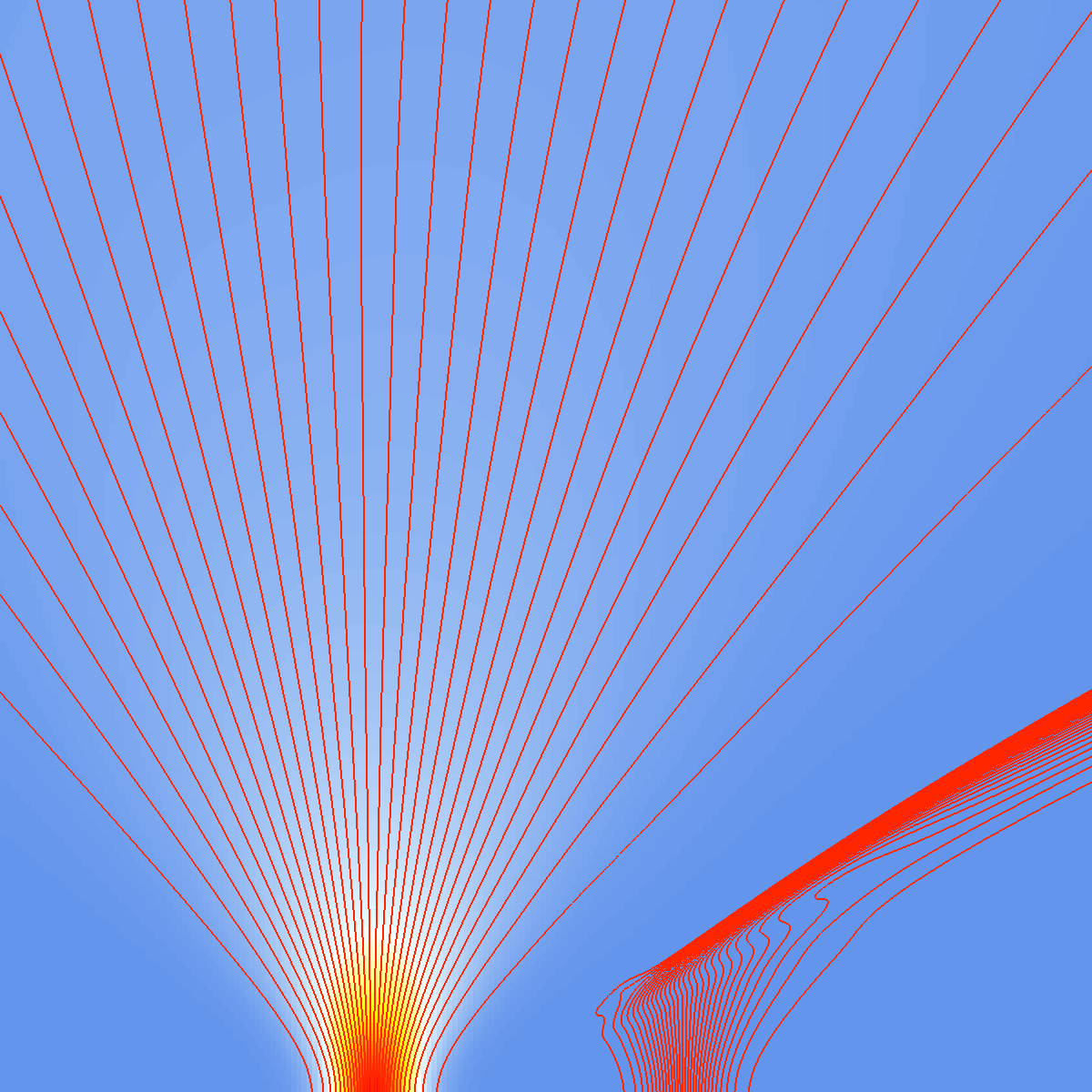}\protect
\par\end{centering}

}\,\subfloat[$a=10^{-10}$\label{fig:sw.2c}]{\protect\begin{centering}
\protect\includegraphics[width=0.31\textwidth]{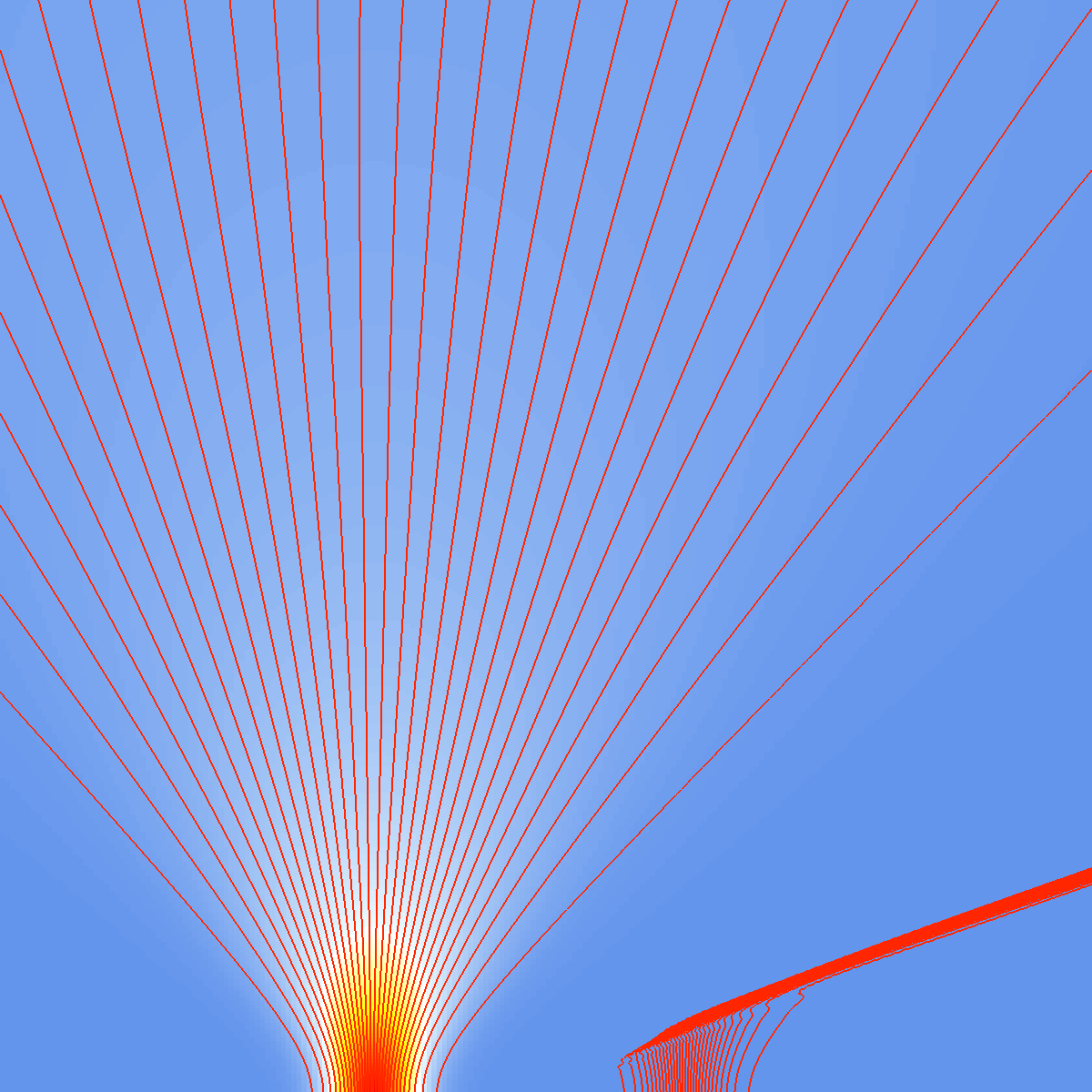}\protect
\par\end{centering}

}
\par\end{centering}

\centering{}\protect\caption{Trajectory behavior during the \textit{quantum sweeper effect} for
different transmission factors $a$ at the right slit of a double
slit setup: With ever lower values of $a$, one can see a steadily
growing tendency for the originally dominant interference fringes
to disappear and for the low counting rate particles of the attenuated
beam to become swept aside. In our model, this phenomenology is explained
by processes of diffusion due to the presence of accumulated kinetic
energy mainly in the ``strong'' beam. The sweeper effect is thus
the result of the vacuum heat sweeping aside the very low intensity
beam, with a ``no crossing'' line defined by the balancing out of
the osmotic momenta coming from the two beams, respectively. Throughout
this paper, to demonstrate the effect more clearly, the same number
of trajectories for each slit is displayed.\label{fig:sw.2}}
\end{figure*}

Interestingly, this is not exactly what one obtains at least for very
low values of $a$ when going through the calculations and/or computer
simulations with our superclassical bouncer model. The latter consists,
among other features, in an explicit form of the velocity field emerging
from the double slit, as well as of the probability density current
associated with it. 

Fig.~\ref{fig:sw.2} shows the \textit{quantum sweeper effect}: a
series of probability density distributions plus averaged trajectories
for the case that the intensity in slit~2 is gradually diminished.
We use the same model as in~\cite{Groessing.2012doubleslit}, or
\cite{Holland.1993}, respectively: particles (represented by plane
waves in the forward $y$-direction) from a coherent source passing
through ``soft-edged'' slits in a barrier (located along the $x$-axis)
and recorded at a screen in the forward direction, i.e.\ parallel
to the barrier. This situation is described by two Gaussians representing
the totality of the effectively ``heated-up'' path excitation field,
one for slit~1 and one for slit~2, whose centers have the same distances
from the plane spanned by the source and the center of the barrier
along the $y$-axis, respectively. Now, with ever lower values of
the transmission factor $a$ during beam attenuation, one can see
a steadily growing tendency for the low counting rate particles of
the attenuated beam to become swept aside. In our model, this is straightforward
to understand, because we have the analytical tools to differentiate
between the forward propagations $\VEC v_{i}$ and the diffusive influences
of velocities $\VEC u_{i}$, as distinguishable contributions from
the different slits~$i$. Thus, it is processes of diffusion which
are seen in operation here, due to the presence of accumulated heat
(i.e.\ kinetic energy), primarily in the ``strong'' beam, as discussed
in the previous Section. So, in effect, we understand Fig.~\ref{fig:sw.2}
as the result of the vacuum heat sweeping aside the very low intensity
beam, with a ``no crossing'' line defined by the balancing out of
the diffusive momenta, $m\left(\VEC u_{1}+\VEC u_{2}\right)=0.$ 

Importantly, for certain slit configurations and sizes of the transmission
factor, the sweeper effect leads to a bunching of trajectories which
may become deflected into a direction almost orthogonal to the original
forward direction. In other words, one would need much wider screens
in the forward direction to register them, albeit then weakened due
to a long traveling distance. On the other hand, if one installed
a screen orthogonal to the ``forward screen'', i.e.\ one that is
parallel to the original forward motion (and thus to the $y$-axis),
one could significantly improve the contrast and thus register the
effect more clearly. Further, we note that changing the distance between
the two slits does not alter the effect, but demonstrates the bunching
of the low counting rate arrivals in essentially the same narrow spatial
area even more drastically. So, again, if one places a screen not
in the forward direction parallel to the barrier containing the double
slit, but orthogonally to the latter, one registers an increased local
density of particle arrivals in a narrow spatial area under an angle
that is independent of the slit distance.

\begin{figure}
\subfloat[$a=1$\label{fig:sw.3a}]{\protect\centering{}\protect\includegraphics[width=0.48\textwidth]{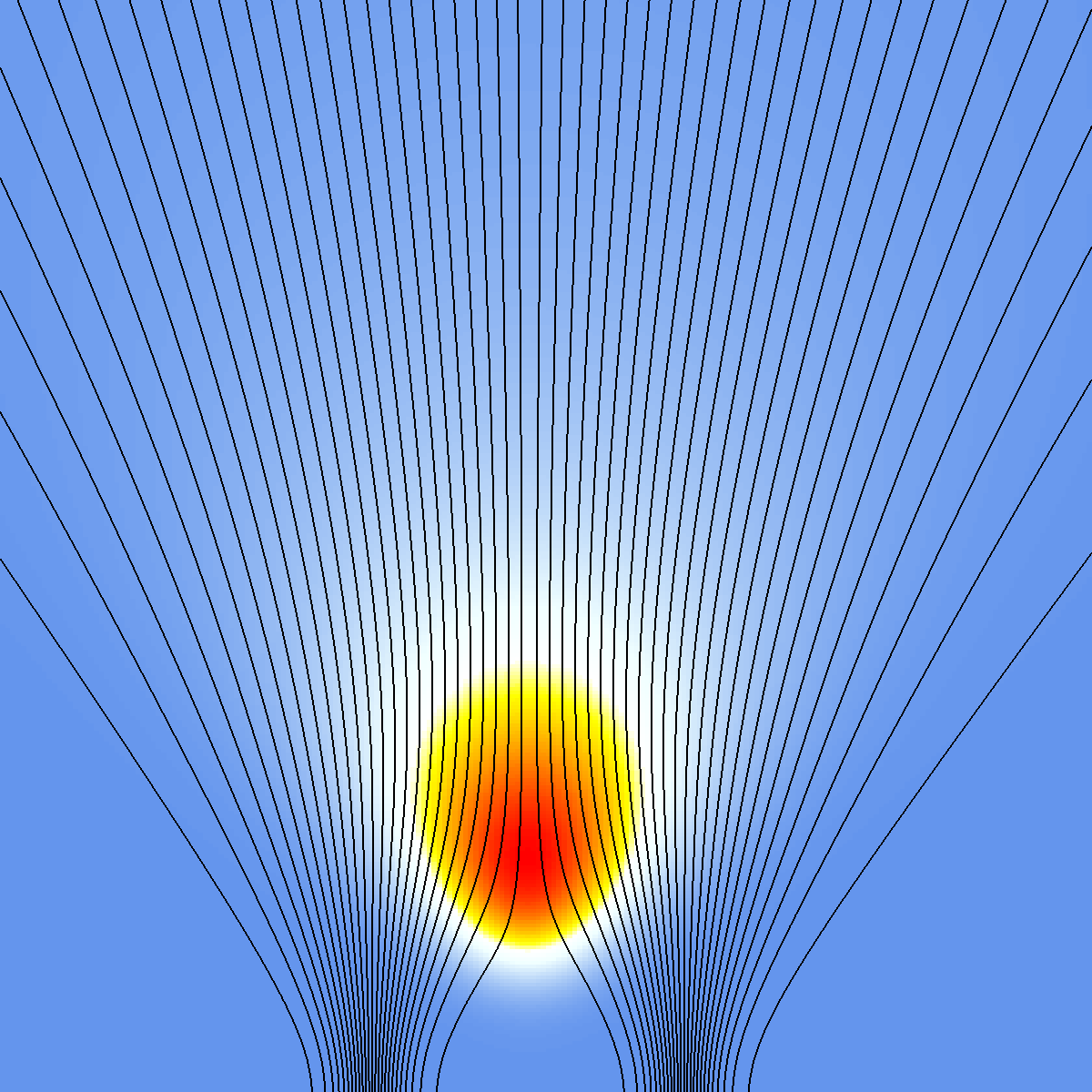}\protect}\hfill{}\subfloat[$a=10^{-8}$\label{fig:sw.3b}]{\protect\centering{}\protect\includegraphics[width=0.48\textwidth]{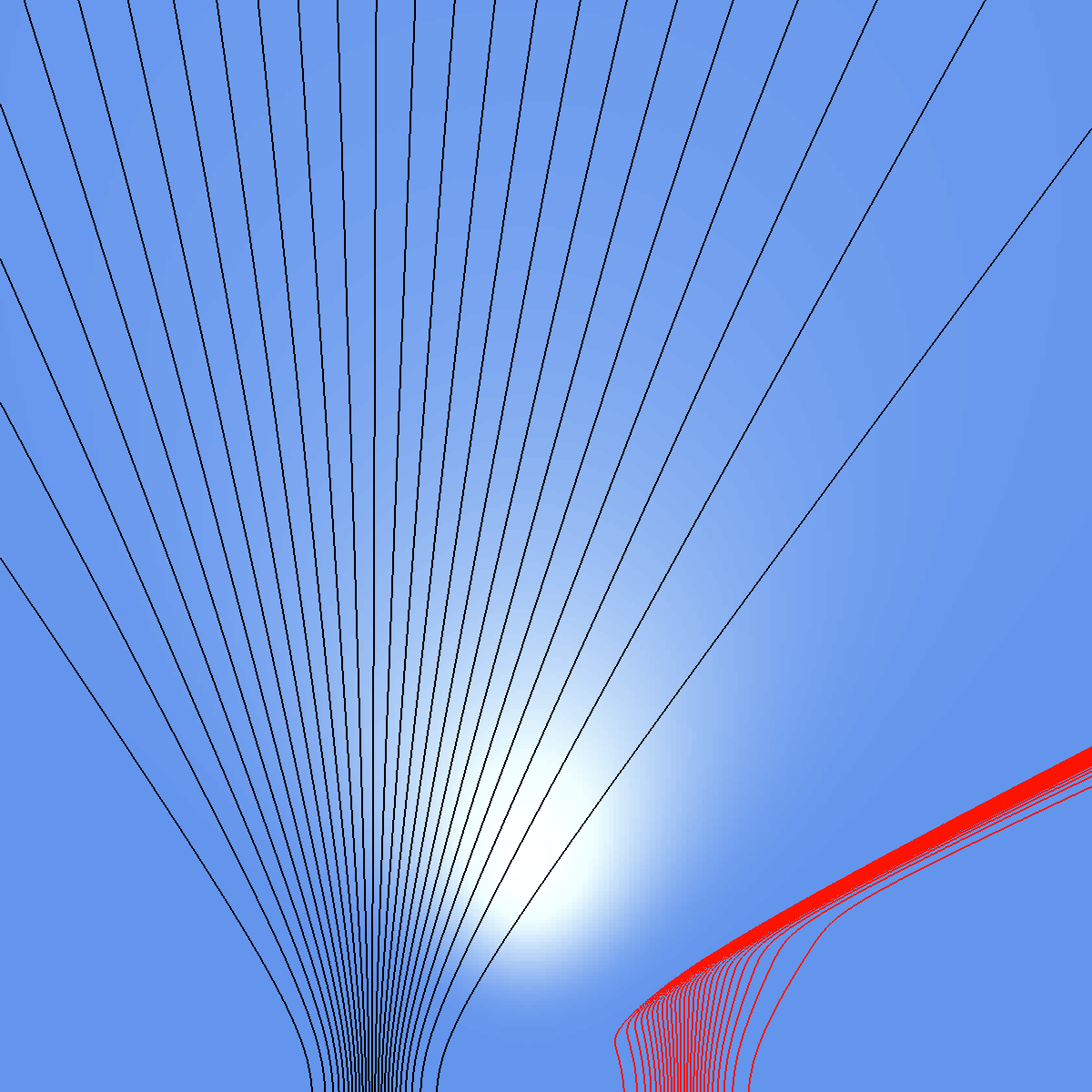}\protect}

\protect\caption{Double-slit experiment with completely incoherent channels. (a) The
average trajectories never cross the central symmetry line, a fact
due to the diffusion related ``hot spot'' indicated in red-to-yellow-to-white
(depicting both interference terms of the total current (\ref{eq:sw.10})),
which represents a kinetic energy reservoir that effectively gives
particles a push in the forward direction. (b) The current's interference
term is now weakened by the factor $a=10^{-8}$, which is why it does
not affect the ``strong'' beam. However, it is sufficient for the
attenuated beam to become deflected and thus to produce the sweeper
effect.\label{fig:sw.3}}
\end{figure}

Let us now turn to the case of decoherent beams. For, although we
shall refrain from constructing a concrete model of decoherence and
implementing it in our scheme, we already have the tools of an effective
theory, i.e.\ to describe decoherence without the need of a specified
mechanism for it. Namely, as full decoherence between two (Gaussian
or other) beams is characterized by the complete absence of the interference
term in the overall probability distribution of the system, this means
that $P_{\mathrm{tot}}=R_{1}^{2}+R_{2}^{2}$, since the interference
term
\begin{equation}
R_{1}R_{2}\left(\VEC v_{\mathrm{1}}+\VEC v_{2}\right)\cos\varphi=0.\label{eq:sw.15}
\end{equation}
If we therefore choose that on average one has $\cos\varphi=0$, a
situation with $\varphi=\frac{\pi}{2}$ effectively describes two
incoherent beams in the double-slit system. What about the two interference
terms in the probability density current~(\ref{eq:sw.10}), then?
Well, the first term is identical with the vanishing (\ref{eq:sw.15}),
but the second term, with $\VEC u_{i}=-\frac{\hbar}{m}$$\frac{\nabla R_{i}}{R_{i}}$
and $\varphi=\frac{\pi}{2}$ explicitly reads as
\begin{equation}
\frac{\hbar}{m}R_{1}R_{2}\left(\frac{\nabla R_{2}}{R_{2}}-\frac{\nabla R_{1}}{R_{1}}\right)=\frac{\hbar}{m}\left(R_{1}\nabla R_{2}-R_{2}\nabla R_{1}\right).\label{eq:sw.16}
\end{equation}
As the distributions $R_{i}$ may have long wiggly tails -- summing
up, after many identical runs, to a Gaussian with no cutoff, but spreading
throughout the whole domain of the experimental setup~\cite{Groessing.2013dice}
--, the expression~(\ref{eq:sw.16}) is not at all guaranteed to
vanish. In fact, a look at Fig.~\ref{fig:sw.3} shows that there
is an effect even for incoherent beams: Although the product $R_{1}R_{2}$
is negligible and therefore leads to no interference fringes on the
screen, nevertheless expression~(\ref{eq:sw.16}) has the effect
of ``bending'' average trajectories so as to obey the ``no crossing''
rule well known from our model as well as from Bohmian theory.

In sum, then, performing a double-slit experiment with decoherent
beams leads to an emergent behavior of particle propagation which
can be explained by the effectiveness of diffusion waves with velocities
$u_{i}$ interacting with each other, thereby creating a ``hot spot''
where the intensity of the diffusive currents is highest and leads
to a deflection into the forward direction such that no crossing of
the average velocities beyond the symmetry line is made possible (Fig.~\ref{fig:sw.3a}).
This is therefore in clear contradiction to the scenario where only
one slit is open for the particle to go through. If the slits are
not open simultaneously, the particles could propagate to locations
beyond the symmetry line, i.e.\ to locations forbidden in the case
of the second slit being open~\cite{Sanz.2009context}.

As our velocity fields $\VEC v_{i}$ and $\VEC u_{i}$~(\ref{eq:sw.12})
are identical with the Bohmian and the ``osmotic'' momentum, respectively,
one can relate them also to the technique of weak measurements. The
latter have turned out~\cite{Leavens.2005weak,Wiseman.2007grounding,Hiley.2012weak}
to provide said velocities as ``weak values'', which are just given
by the real and complex parts of the quantum mechanical expression
$\frac{\left\langle \mathbf{r}\mid\hat{p}\mid\varPsi\left(\mathbf{\mathit{t}}\right)\right\rangle }{\left\langle \mathbf{r}\mid\varPsi\left(\mathbf{\mathit{t}}\right)\right\rangle }$,
i.e.\ the weak values associated with a weak measurement of the momentum
operator $\hat{p}$ followed by the usual (``strong'') measurement
of the position operator $\hat{r}$ whose outcome is $\mathbf{r}$.
In other words, in principle the trajectories for intensity hybrids
generally, and for the quantum sweeper in particular, are therefore
accessible to experimental confirmation.

\section{The quantum mechanical description of the sweeper effect\label{sec:sweeper quantum}}

\begin{figure}
\begin{centering}
\subfloat[Probability density distribution $P$ in a distance of 5\,m from
the double slit. The two cases of the attenuation factor at the right
slit of a double slit system, i.e. $a=10^{-4}$ and $a=10^{-8}$,
respectively, essentially provide the same distribution at moderate
resolution. \label{fig:gupferl-a}]{\protect\begin{centering}
\protect\includegraphics{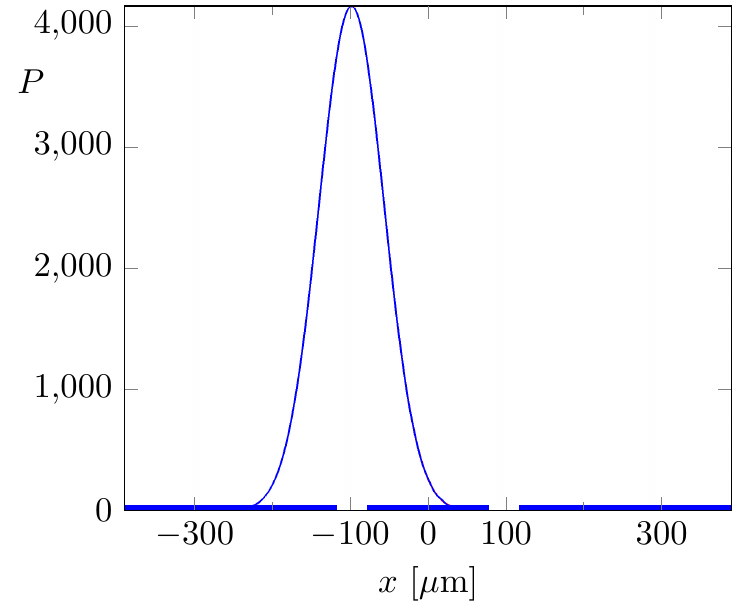}\protect
\par\end{centering}

}\hfill{}\subfloat[Same as in (a), but zooming in with a factor of 1,000. Now the two
cases are discernible: note the faint rest of interference phenomena
for the case of $a=10^{-4}$ (blue), whereas for $a=10^{-8}$ (red)
apparently smooth behavior is seen. \label{fig:gupferl-b}]{\protect\begin{centering}
\protect\includegraphics{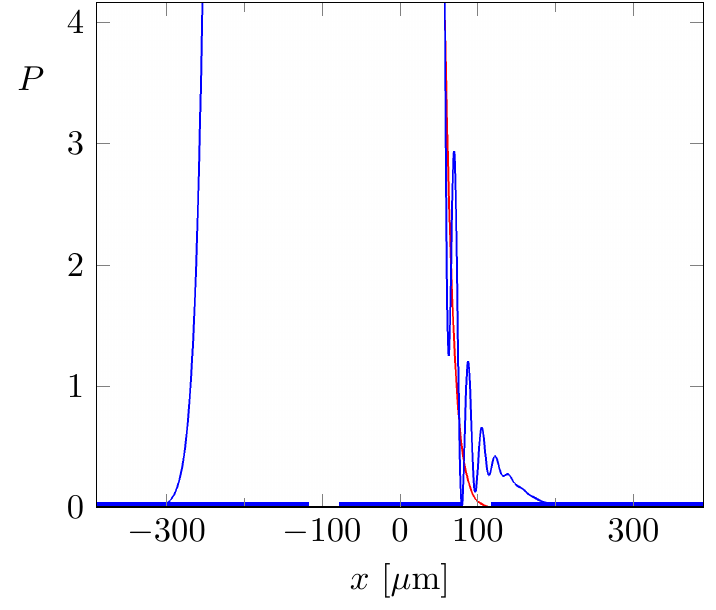}\protect
\par\end{centering}

}
\par\end{centering}

\begin{centering}
\medskip{}
\subfloat[Same as in (a), but now on a logarithmic scale. Dotted initial distributions
for the cases of $a=10^{-4}$ (blue) and $a=10^{-8}$ (red), respectively,
develop into distributions clearly showing interference phenomena
which have been ``swept aside'' far to the right. This sweeper effect
is due to the explicit appearance of the nonlinear structure of the
probability density current in these domains for very low values of
$a$. \label{fig:gupferl-c}]{\protect\begin{centering}
\protect\includegraphics{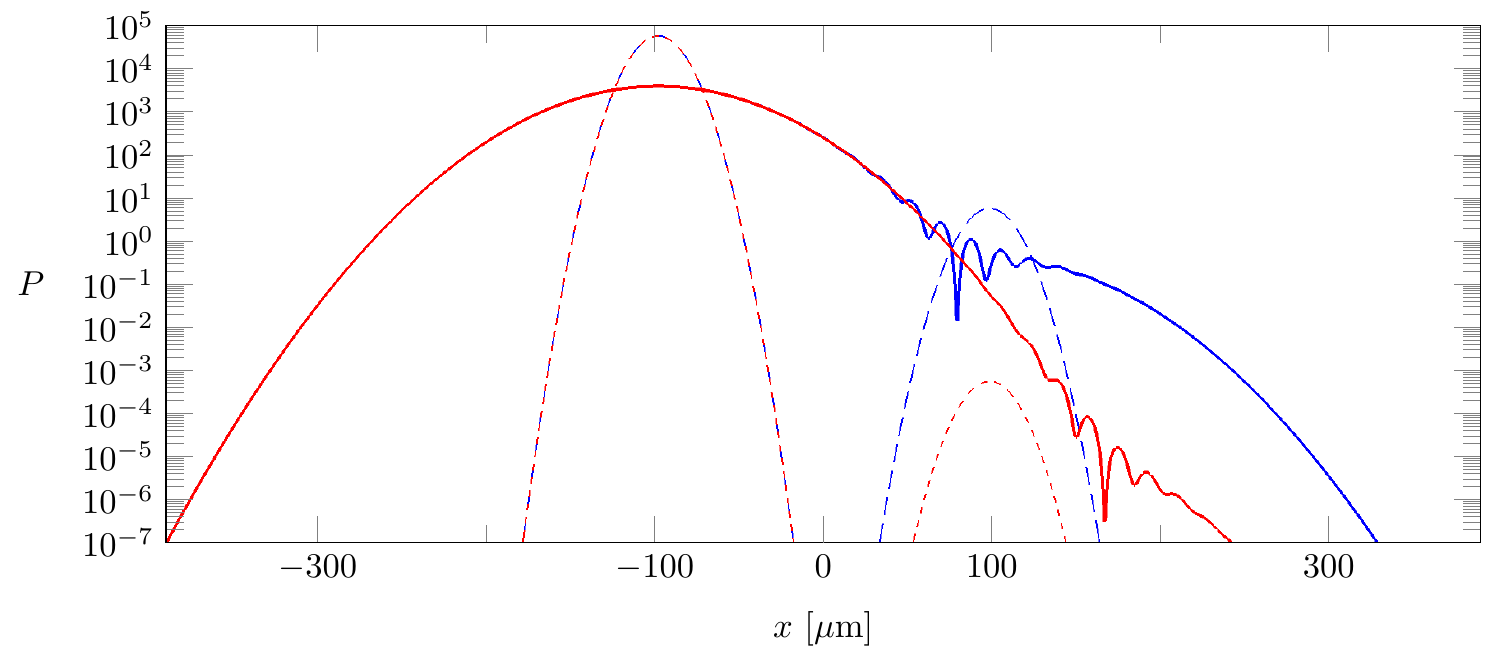}\protect
\par\end{centering}

}
\par\end{centering}

\centering{}\protect\caption{The sweeper effect as described by quantum mechanics\label{fig:gupferl}}
\end{figure}
Let us now consider the stochastic attenuation discussed above in
purely quantum mechanical terms. As already mentioned, the probability
density distribution is given by Equation (\ref{eq:sw.2}). A graphic
representation of this distribution in a distance of 5m from the double
slit is shown in Fig.~\ref{fig:gupferl}. Two cases of the attenuation
factor at one of the two slits of a double slit system are shown,
i.e. $a=10^{-4}$ and $a=10^{-8}$ affecting the right slit, respectively.
As is to be expected, on a linear scale the distribution will appear
as if practically the whole intensity goes through the left un-attenuated
slit (Fig.~\ref{fig:gupferl-a}). Zooming in with a factor of 1,000
as shown in Fig.~\ref{fig:gupferl-b}, one can see the faint rest
of interference phenomena for the case of $a=10^{-4}$ (blue), whereas
for $a=10^{-8}$ (red) apparently smooth behavior is seen. Still,
the full effect is best visible on the logarithmic scale shown in
Fig.~\ref{fig:gupferl-c}. Compared to the dotted initial distributions
for the cases of $a=10^{-4}$ (blue) and $a=10^{-8}$ (red), respectively,
the whole distribution clearly shows interference phenomena which
have been ``swept aside'' far to the right. Thus, the quantum sweeper
effect is confirmed also via orthodox language. 

The bunching together of low counting rate particles within a very
narrow spatial domain, or channel, respectively, counters naive expectations
that with ever higher beam attenuation nothing interesting may be
seen any more. The reason why these expectations are not met is given
by the explicit appearance of the nonlinear structure of the probability
density current~(\ref{eq:sw.10}) in these domains for very low values
of $a$.

\section{Summary and implications\label{sec:conclusion}}

As for implications of our finding presented here, we shall now briefly
remark on its relevance with respect to the issue of wave-particle
duality. Considering the appearance of compressed interference fringes
in the attenuated beam in Fig.~\ref{fig:sw.2}, it is indisputable
that one has to do with the result of a wave-like behavior. This is
confirmed in Fig.~\ref{fig:sw.3b} where the decoherent scenario
is characterized by the complete absence of such wave-like behavior
like interference fringes. This means, however, that an often used
argument to describe the complementarity between wave- and particle-like
behavior in the double slit experiment, or in interferometry, respectively,
has only limited applicability, as it does not apply to intensity
hybrids, since in our model the wave-like contributions due to diffusion
are always present. Specifically, the relation for pure states~\cite{Greenberger.1988simultaneous}
\begin{equation}
D^{2}+V^{2}=1,
\end{equation}
with distinguishability 
\begin{equation}
D=\left|\frac{R_{1}^{2}-R_{2}^{2}}{R_{1}^{2}+R_{2}^{2}}\right|
\end{equation}
 representing the particle-related which-path information and visibility
\begin{equation}
V=\frac{\mid R_{1}+R_{2}\mid^{2}-\mid R_{1}-R_{2}\mid^{2}}{\mid R_{1}+R_{2}\mid^{2}+\mid R_{1}-R_{2}\mid^{2}}=\frac{2R_{1}R_{2}}{R_{1}^{2}+R_{2}^{2}}
\end{equation}
the contrast of the interference fringes in the standard quantum mechanical
double slit scenario, suggests that with ever lower values of $a$
ever lower values of $V$ are implied in a constantly decreasing manner.
In our case, by considering the superclassical nature of the sweeper
effect, we find a deviating, characteristic signature at very low
values of $a\lesssim10^{-4}$. In this domain, the usual expectation
would be that practically one has arrived at the ``particle'' side
of the complementarity principle, i.e.\ essentially a one-slit distribution,
with wave-like phenomena having almost disappeared. However, if one
has a very strongly attenuated beam, the emerging behavior of its
outgoing trajectories is different from a one-slit particle distribution
scenario if the other slit is open and un-attenuated. The increased
local relative contrast corresponding to a bunching of trajectories
and due to said nonlinear effect of the probability density current
is to be captured by a vertical screen (i.e.\ parallel to the $y$-axis)
for optimal visibility.

We have thus shown that for transmission factors below $a\lesssim10^{-4}$
in intensity hybrids, new effects appear which are not taken into
account in a naive, i.e.\ linear, extrapolation of expectations based
on higher-valued transmission factors. We have described the phenomenology
of these \textit{quantum sweeper} effects, including the bunching
together of low counting rate particles within a very narrow spatial
domain, or channel, respectively. However, we also stress that these
results are in accordance with standard quantum mechanics, since we
just used a re-labeling and re-drawing of the constituent parts of
the usual quantum mechanical probability density currents. However,
concerning the explicit phenomenological appearances due to the nonlinear
structure of the probability density current in the respective domains
for very low values of $a$, our subquantum model is better equipped
to deal with these appearances explicitly.

With the discovery of the quantum sweeper effect on the basis of a
superclassical causal approach to quantum mechanics, we claim to have
presented a first example as it was demanded by Rabi. We are optimistic
that through further developments, both in superclassical theory employing
subquantum mechanics and in weak measurement techniques capable of
probing the latter regime, more unexpected new effects can be predicted
and eventually be confirmed in experiment. 
\begin{acknowledgments}
We thank Jan Walleczek for many enlightening discussions, and the
Fetzer Franklin Fund for partial support of the current work.
\end{acknowledgments}


\begin{thebibliography}{10}

\bibitem{FreireJr..2005science}
O.~Freire~Jr., ``Science and exile: {D}avid {B}ohm, the cold war, and a new
  interpretation of quantum mechanics,''
  \href{http://dx.doi.org/10.1525/hsps.2005.36.1.1}{{\em {HSPS}} {\bfseries 36}
  (2005) 1--34}, \href{http://arxiv.org/abs/physics/0508184}{{\ttfamily
  arXiv:physics/0508184 [physics.hist-ph]}}.

\bibitem{Groessing.2014attenuation}
G.~Gr{\"o}ssing, S.~Fussy, J.~Mesa~Pascasio, and H.~Schwabl, ``Extreme beam
  attenuation in double-slit experiments: {Q}uantum and subquantum scenarios,''
  \href{http://dx.doi.org/10.1016/j.aop.2014.11.015}{{\em Ann. Phys.}
  {\bfseries 353} (2015) 271--281},
  \href{http://arxiv.org/abs/1406.1346}{{\ttfamily arXiv:1406.1346
  [quant-ph]}}.

\bibitem{Groessing.2015implications}
G.~Gr{\"o}ssing, S.~Fussy, J.~Mesa~Pascasio, and H.~Schwabl, ``Implications of
  a deeper level explanation of the {deBroglie--Bohm} version of quantum
  mechanics,'' \href{http://dx.doi.org/10.1007/s40509-015-0031-0}{{\em Quantum
  Stud.: Math. Found.} (2015) },
  \href{http://arxiv.org/abs/1412.8349}{{\ttfamily arXiv:1412.8349
  [quant-ph]}}. In Press.

\bibitem{Rauch.1984static}
H.~Rauch and J.~Summhammer, ``Static versus time-dependent absorption in
  neutron interferometry,''
  \href{http://dx.doi.org/10.1016/0375-9601(84)90586-3}{{\em Phys. Lett. A}
  {\bfseries 104} (1984) 44--46}.

\bibitem{Rauch.1990low-contrast}
H.~Rauch, J.~Summhammer, M.~Zawisky, and E.~Jericha, ``Low-contrast and
  low-counting-rate measurements in neutron interferometry,''
  \href{http://dx.doi.org/10.1103/PhysRevA.42.3726}{{\em Phys. Rev. A}
  {\bfseries 42} (1990) 3726--3732}.

\bibitem{Groessing.2012emerqum11-book}
G.~Gr{\"o}ssing, ed., {\em Emergent Quantum Mechanics 2011}.
\newblock No.~361/1. {IOP} Publishing, Bristol, 2012.
\newblock Url: http://iopscience.iop.org/1742-6596/361/1.

\bibitem{Groessing.2014emqm13-book}
G.~Gr{\"o}ssing, H.-T. Elze, J.~Mesa~Pascasio, and J.~Walleczek, eds., {\em
  Emergent Quantum Mechanics 2013}.
\newblock No.~504/1. {IOP} Publishing, Bristol, 2014.
\newblock Url: http://iopscience.iop.org/1742-6596/504/1.

\bibitem{Fort.2010path-memory}
E.~Fort, A.~Eddi, A.~Boudaoud, J.~Moukhtar, and Y.~Couder, ``Path-memory
  induced quantization of classical orbits,''
  \href{http://dx.doi.org/10.1073/pnas.1007386107}{{\em {PNAS}} {\bfseries 107}
  (2010) 17515--17520}.

\bibitem{Couder.2006single-particle}
Y.~Couder and E.~Fort, ``Single-particle diffraction and interference at a
  macroscopic scale,''
  \href{http://dx.doi.org/10.1103/PhysRevLett.97.154101}{{\em Phys. Rev. Lett.}
  {\bfseries 97} (2006) 154101}.

\bibitem{Couder.2012probabilities}
Y.~Couder and E.~Fort, ``Probabilities and trajectories in a classical
  wave-particle duality,''
  \href{http://dx.doi.org/10.1088/1742-6596/361/1/012001}{{\em J. Phys.: Conf.
  Ser.} {\bfseries 361} (2012) 012001}.

\bibitem{Groessing.2010emergence}
G.~Gr{\"o}ssing, S.~Fussy, J.~Mesa~Pascasio, and H.~Schwabl, ``Emergence and
  collapse of quantum mechanical superposition: {O}rthogonality of reversible
  dynamics and irreversible diffusion,''
  \href{http://dx.doi.org/10.1016/j.physa.2010.07.017}{{\em Physica A}
  {\bfseries 389} (2010) 4473--4484},
  \href{http://arxiv.org/abs/1004.4596}{{\ttfamily arXiv:1004.4596
  [quant-ph]}}.

\bibitem{Fussy.2014multislit}
S.~Fussy, J.~Mesa~Pascasio, H.~Schwabl, and G.~Gr{\"o}ssing, ``Born's rule as
  signature of a superclassical current algebra,''
  \href{http://dx.doi.org/10.1016/j.aop.2014.02.002}{{\em Ann. Phys.}
  {\bfseries 343} (2014) 200--214},
  \href{http://arxiv.org/abs/1308.5924}{{\ttfamily arXiv:1308.5924
  [quant-ph]}}.

\bibitem{Groessing.2012doubleslit}
G.~Gr{\"o}ssing, S.~Fussy, J.~Mesa~Pascasio, and H.~Schwabl, ``An explanation
  of interference effects in the double slit experiment: {C}lassical
  trajectories plus ballistic diffusion caused by zero-point fluctuations,''
  \href{http://dx.doi.org/10.1016/j.aop.2011.11.010}{{\em Ann. Phys.}
  {\bfseries 327} (2012) 421--437},
  \href{http://arxiv.org/abs/1106.5994}{{\ttfamily arXiv:1106.5994
  [quant-ph]}}.

\bibitem{Sanz.2008trajectory}
A.~S. Sanz and S.~Miret-Art{\'e}s, ``A trajectory-based understanding of
  quantum interference,''
  \href{http://dx.doi.org/10.1088/1751-8113/41/43/435303}{{\em J. Phys. A:
  Math. Gen.} {\bfseries 41} (2008) 435303},
  \href{http://arxiv.org/abs/0806.2105}{{\ttfamily arXiv:0806.2105
  [quant-ph]}}.

\bibitem{Holland.1993}
P.~R. Holland, {\em The Quantum Theory of Motion}.
\newblock Cambridge University Press, Cambridge, {UK}, 1993.

\bibitem{Groessing.2013dice}
G.~Gr{\"o}ssing, S.~Fussy, J.~Mesa~Pascasio, and H.~Schwabl, ``{'Systemic}
  nonlocality' from changing constraints on sub-quantum kinematics,''
  \href{http://dx.doi.org/10.1088/1742-6596/442/1/012012}{{\em J. Phys.: Conf.
  Ser.} {\bfseries 442} (2013) 012012},
  \href{http://arxiv.org/abs/1303.2867}{{\ttfamily arXiv:1303.2867
  [quant-ph]}}.

\bibitem{Sanz.2009context}
A.~S. Sanz and F.~Borondo, ``Contextuality, decoherence and quantum
  trajectories,'' \href{http://dx.doi.org/10.1016/j.cplett.2009.07.061}{{\em
  Chem. Phys. Lett.} {\bfseries 478} (2009) 301--306},
  \href{http://arxiv.org/abs/0803.2581}{{\ttfamily arXiv:0803.2581
  [quant-ph]}}.

\bibitem{Leavens.2005weak}
C.~R. Leavens, ``Weak measurements from the point of view of {B}ohmian
  mechanics,'' \href{http://dx.doi.org/10.1007/s10701-004-1984-8}{{\em Found.
  Phys.} {\bfseries 35} (2005) 469--491}.

\bibitem{Wiseman.2007grounding}
H.~M. Wiseman, ``Grounding {B}ohmian mechanics in weak values and
  {b}ayesianism,'' \href{http://dx.doi.org/10.1088/1367-2630/9/6/165}{{\em New
  J. Phys.} {\bfseries 9} (2007) 165---176},
  \href{http://arxiv.org/abs/0706.2522}{{\ttfamily arXiv:0706.2522
  [quant-ph]}}.

\bibitem{Hiley.2012weak}
B.~J. Hiley, ``Weak values: {A}pproach through the {C}lifford and {M}oyal
  algebras,'' \href{http://dx.doi.org/10.1088/1742-6596/361/1/012014}{{\em J.
  Phys.: Conf. Ser.} {\bfseries 361} (2012) 012014},
  \href{http://arxiv.org/abs/1111.6536v1}{{\ttfamily arXiv:1111.6536v1
  [quant-ph]}}.

\bibitem{Greenberger.1988simultaneous}
D.~M. Greenberger and A.~Yasin, ``Simultaneous wave and particle knowledge in a
  neutron interferometer,''
  \href{http://dx.doi.org/10.1016/0375-9601(88)90114-4}{{\em Phys. Lett. A}
  {\bfseries 128} (1988) 391--394}.

\end{thebibliography}
\providecommand{\href}[2]{#2}\begingroup\raggedright\endgroup

\end{document}